\documentclass[11pt,a4paper]{JINST}
\pdfoutput=1
\usepackage{amsmath}
\usepackage{amsfonts}

\newcommand{\mbf}[1]{\ensuremath{\mathbf{#1}}}
\newcommand{\mrm}[1]{\ensuremath{\mathrm{#1}}}

\title{Robustness of Planar Fourier Capture Arrays to Colour Changes and Lost Pixels}

\author{Patrick R. Gill$^{ab}$\thanks{Corresponding
author.}~, Changhyuk Lee$^a$, Sriram Sivaramakrishnan$^a$ and Alyosha Molnar$^a$\\
\llap{$^a$}School of Electrical and Computer Engineering, Cornell University\\
223 Phillips Hall, Ithaca, NY 14853\\
USA\\
\llap{$^b$}Department of Psychology, University of Toronto\\
100 St. George Street, Toronto, ON M5S 3G3, Canada\\
\\
E-mail: \email{patrick.gill@utoronto.ca}}

\abstract{Planar Fourier capture arrays (PFCAs) are optical sensors built entirely in standard microchip manufacturing flows.  PFCAs are composed of ensembles of angle sensitive pixels (ASPs) that each report a single coefficient of the Fourier transform of the far-away scene.  Here we characterize the performance of PFCAs under the following three non-optimal conditions.  First, we show that PFCAs can operate while sensing light of a wavelength other than the design point. Second, if only a randomly-selected subset of 10\% of the ASPs are functional, we can nonetheless reconstruct the entire far-away scene using compressed sensing.  Third, if the wavelength of the imaged light is unknown, it can be inferred by demanding self-consistency of the outputs.}


\begin{document}

\section{Introduction}

\subsection{Planar Fourier Capture Arrays}

Planar Fourier capture arrays (PFCAs)  \cite{PFCA2011} are ensembles of angle sensitive pixels (ASPs)  \cite{wang2009light,Sensors2010Poster} that directly capture the Fourier transform of a far-away image.  The light sensitivity of an individual ASP is a sinusoidal function of incident angle  \cite{wang2009light}.  A carefully-chosen ensemble of ASPs can report the entire Fourier transform of the far-away scene up to the Nyquist limit set by the highest-frequency ASP in the ensemble  \cite{PFCA2011}.  As ASPs are extremely thin, light and cheap to manufacture, PFCAs have the potential to become a disruptive technology in the fields of sensing and robotics  \cite{batchelor1985automated,lillesand2004remote,urmson2008autonomous,bergstrom2009integration} when volume, mass or cost constraints are paramount.  

For PFCAs to reach their full potential, it would be useful to understand some of the limitations and freedoms afforded by this new class of imager.  Our first PFCA was optimized for green incident light, and the reconstruction algorithms we used previously  \cite{PFCA2011} assume all sensors are functional.  In this publication we thoroughly report PFCA design considerations, then investigate the performance of PFCAs imaging light of a different colour than the design specification, both when that colour is known {\em a priori} and when it must be determined blindly.  Further, we will show that by using a signal-processing technique called compressed sensing, we can reconstruct images even when the outputs of only a small fraction of the ASPs are available.  

\subsection{Angle Sensitive Pixels}

\begin{figure} 
\centerline{\includegraphics[width=.6\textwidth]{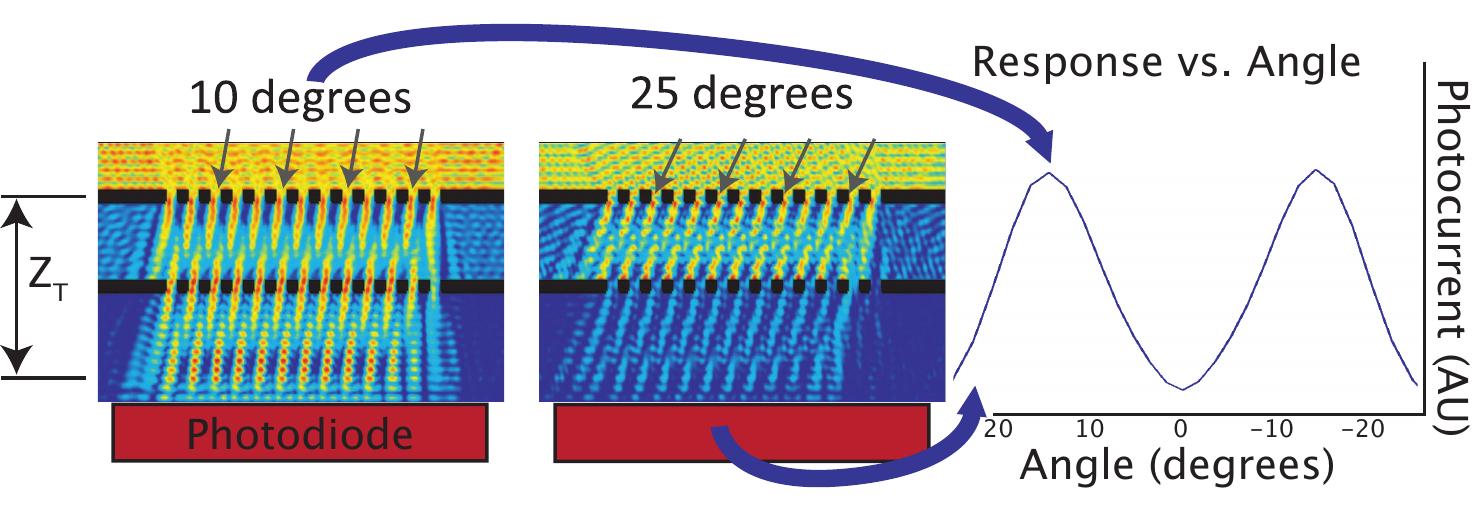}}
\caption{Interaction Between Incident Light and ASPs.  Left: light incident at $10\,^{\circ}$ produces spatial maxima that align with gaps in the second grating, resulting in a large photocurrent.  Center: light incident at $25\,^{\circ}$ produces spatial maxima that are largely blocked by the second grating, resulting in a small photocurrent.  Right: photocurrent is a sinusoidal function of incident angle.} 
\label{ASPfig} 
\end{figure}

Angle sensitive pixels (ASPs)  \cite{wang2009light,Sensors2010Poster} are photosensitive pixels that, using diffraction gratings, couple to far-field light sources with an efficiency that is a sinusoidal function of incident angle.  ASPs attain their angle sensitivity through the moir\'{e} effect using two metal gratings with identical spacing but at different heights above a photodiode below (see Figure \ref{ASPfig}).  Due to the vertical displacement of the gratings, the angle subtended by one spatial period of the top grating viewed by the photodiode is slightly less than that subtended by the bottom grating.  This angular disparity invokes the moir\'{e} effect, making the effective transmission aperture a sinusoidal function of incident angle.  Both of the gratings can be manufactured using metal interconnect layers intrinsic to the CMOS process, and the photodiodes can be manufactured using intrinsic semiconductors, meaning ASPs can be manufactured entirely using existing unmodified CMOS process flows.  

\begin{figure} 
\centerline{\includegraphics[width=.6\textwidth]{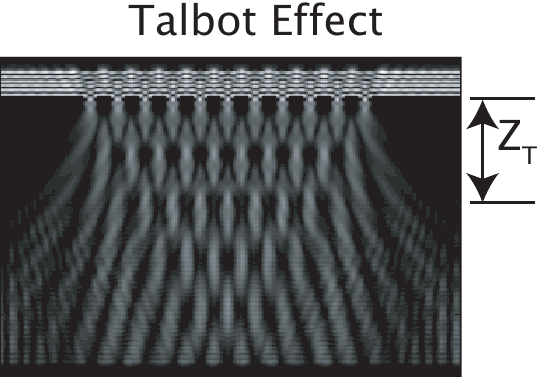}} 
\caption{The Talbot Effect.  
Wave simulation of monochromatic light striking a grating at the normal.  The grating periodicity is recapitulated at regularly-spaced depths below the grating; a phenomenon known as the Talbot effect.  The Talbot depth $Z_T$ is shown at right.} 
\label{MTfig} 
\end{figure}

The moir\'{e} effect is essentially a ray optics effect, however at the small spatial scale of CMOS structures photons cannot be modeled as non-diffracting rays.  Fortunately, the moir\'{e} effect can be rescued by exploiting the Talbot effect  \cite{talbot1836lxxvi,teng2008quasi}.  The Talbot effect is the property that a regular periodic grating illuminated by monochromatic light creates self-images at integer multiples of the Talbot depth $Z_T = \frac{2p^2}{\lambda} - \frac{\lambda}{2}$ (see Section \ref{halfTalSec}) where $p$ is the grating pitch (spatial period) and $\lambda$ is the wavelength of light (see Figures \ref{ASPfig} and \ref{MTfig}).  By ensuring that the second grating is located at an integer multiple of half\footnote{At half-integer multiples of the Talbot depth the periodicity of the initial grating reemerges, but with its phase reversed; see Figures \ref{ASPfig} and \ref{MTfig}.} the Talbot depth, the moir\'{e} effect is rescued and further diffraction effects can be neglected. 
 
The transfer function of light incident on an ASP can be modeled as in Equation \ref{lightPix}.

  \begin{equation}
  \label{lightPix}
R=I_0(1-m \cos(b \theta + \alpha))F(\theta)(1 + \eta),    
  \end{equation}
  where $R$ is the readout of the ASP, $I_0$ is proportional to the light flux at the ASP, $\theta$ is the incident angle along the sensitive axis, $b$ is the angular sensitivity (designed to range from 7 to 39 - see Equation \ref{betaEq}), $m$ is the modulation depth of the ASP (see Equation \ref{talDep}), $\alpha$ is a designable phase offset caused by a displacement between the top and bottom gratings, $F(\theta)$ is a slowly-varying aperture function and $\eta$ is multiplicative noise.  
\subsection{Deriving the Talbot Depth}
\label{halfTalSec}
\begin{figure} 
\centerline{\includegraphics[width=.3\textwidth]{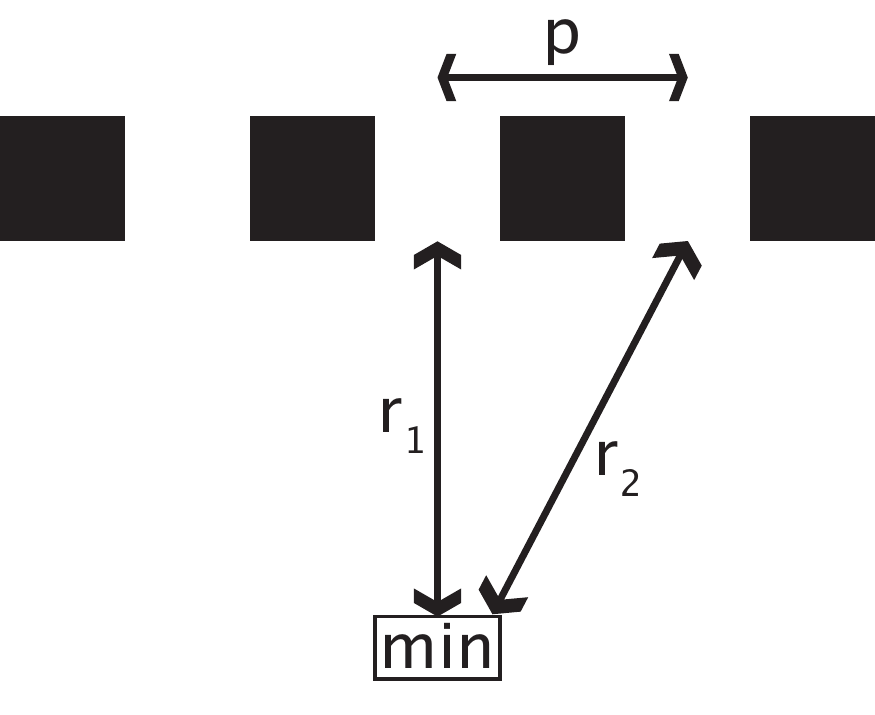}} 
\caption{Calculating the Talbot Depth.  The geometry in this configuration leads to our expression for the half Talbot depth in Equation \protect \ref{talDepHalf}} 
\label{Tquart} 
\end{figure}

The $\frac{\lambda}{2}$ term of $Z_T = \frac{2p^2}{\lambda} - \frac{\lambda}{2}$ is often omitted especially in considerations of Talbot periodicity rather than absolute depth and when $p^2 \gg \lambda^2$, but is derivable from a consideration of optical path lengths and provides a noticeable increase in accuracy when modeling and designing ASPs.  Consider the geometry of Figure \ref{Tquart}.  For a local minimum to occur at $r_1 = \frac{Z_T}{2}$, the path length difference between $r_1 = \frac{Z_T}{2}$ and $r_2 = \sqrt{\left(\frac{Z_T}{2}\right)^2 + p^2}$ should be $\frac{\lambda}{2}$.  Thus,

\begin{eqnarray}
r_2 - r_1 & = & \frac{\lambda}{2}  \nonumber \\ 
\sqrt{\left(\frac{Z_T}{2}\right)^2 + p^2} - \frac{Z_T}{2} & = & \frac{\lambda}{2}  \nonumber \\ 
\sqrt{\left(\frac{Z_T}{2}\right)^2 + p^2} &  = & \frac{Z_T}{2}  + \frac{\lambda}{2}  \nonumber \\
\left(\frac{Z_T}{2}\right)^2 + p^2 &  = & \left(\frac{Z_T}{2}\right)^2 + \frac{Z_T}{2}\lambda + \frac{\lambda^2}{4}  \nonumber \\
p^2 &  = & \frac{Z_T}{2}\lambda + \frac{\lambda^2}{4}  \nonumber \\
\frac{Z_T}{2} & = & \frac{p^2}{\lambda} - \frac{\lambda}{4} \label{talDepHalf}
\end{eqnarray}

\section{Designing PFCAs}

An ASP's angular sensitivity ($b$ from Equation \ref{lightPix}) can be found by considering the angular displacement required for a lobe of the Talbot diffraction pattern to traverse a full period of the analyzer grating, as in Equation \ref{betaEq}.  \begin{equation}
  \label{betaEq}
b = 2 \pi \frac{z_{\operatorname{effective}}}{p n} ,    
  \end{equation} 
where $z_{\operatorname{effective}}$ is the effective vertical displacement between gratings (see Equation \ref{changeA}) 
and $n$ is the refractive index of the medium, in this case SiO$_2$.  The depths $z$ at which modulation depth $m$ is maximal correspond to integer multiples of half the Talbot depth, as in Equation \ref{talDep}.
 \begin{equation}
  \label{talDep}
m\, \operatorname{maximal} \, \operatorname{when} \, z = a \left( \frac{p^2}{\lambda} - \frac{\lambda}{4} \right); \, \, a \in \mathbb{I}.
  \end{equation}

\subsection{Effective Depth}

Since the PFCA was built, detailed measurements have indicated that when using Equation \ref{betaEq}, $b$ is more accurately predicted by replacing the vertical separation between gratings $z$ with an effective depth, $z_{\operatorname{effective}}$, set by the nearest half-Talbot depth as follows.


\begin{eqnarray}
a_{\operatorname{closest\,Talbot}} & = & \operatorname{round} \left( \frac{z}{\left( \frac{p^2}{\lambda} - \frac{\lambda}{4} \right)} \right) \nonumber \\ 
z_{\operatorname{effective}} & = & a_{\operatorname{closest\,Talbot}} \left( \frac{p^2}{\lambda} - \frac{\lambda}{4} \right) \label{changeA}
\end{eqnarray}

At optimal depths for a particular $\lambda$, $z_{\operatorname{effective}} = z$.  However, for non-optimal $\lambda$s, $z_{\operatorname{effective}}$ and thus $b$ vary as a function of $\lambda$, an effect previously observed \cite{PFCA2011,wang2009light,Sensors2010Poster} but not explained.

\begin{figure} 
\centerline{\includegraphics[width=.6\textwidth]{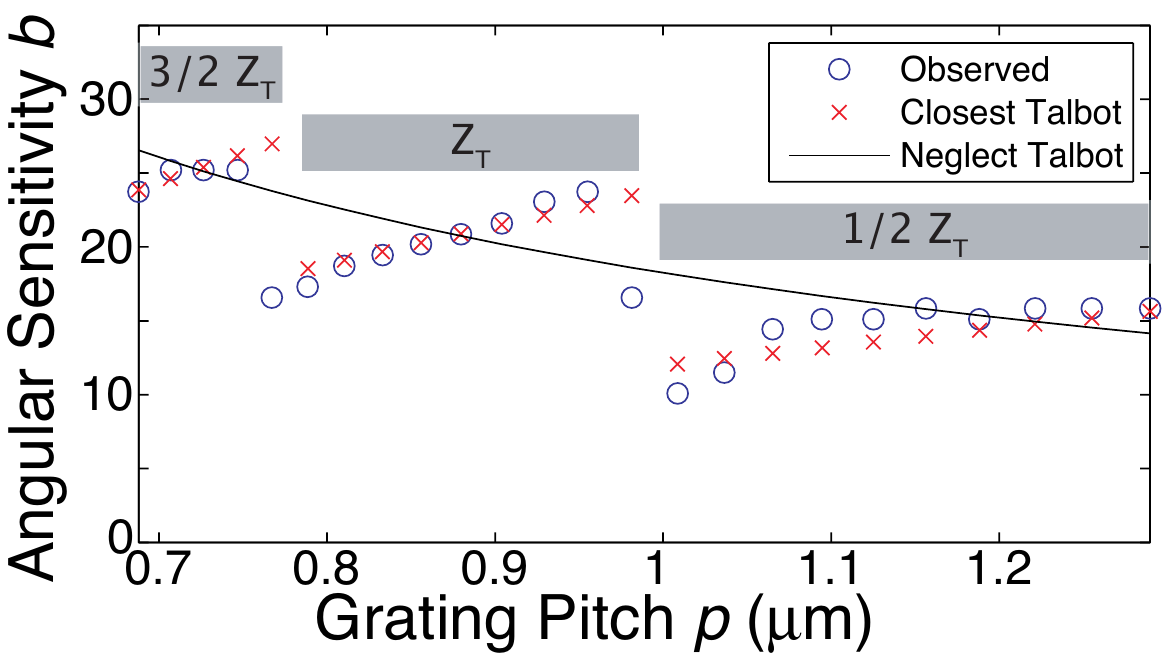}}
\caption{Closest Talbot Pattern Determines $b$.  Observed $b(p)$ agrees better with the assumption that $z_{\operatorname{effective}}$ equals the closest optimal Talbot depth, and not the manufactured depth. Blue circles: measured $b$ as a function of pitch for incident green light on a series of devices spanning a range of $p$ from 0.7 to 1.3 microns.  Red $\mrm{x}$s: modeled $b$ assuming Equations  \protect \ref{betaEq}  and \protect \ref{changeA}.  Black line: modeled $b$ assuming $z_{\operatorname{effective}} = z$ and Equation \protect \ref{betaEq}.  Grey bars: depth of closest Talbot pattern.} 
\label{placeholder} 
\end{figure}

The phenomenon that the effective depth is the closest ideal Talbot pattern was demonstrated in a test structure accompanying our original PFCA prototype.  A long, linear array of photodetectors was built with smoothly varying pitch.  The observed angular sensitivity $b$ as a function of pitch $p$ is plotted in Figure \ref{placeholder}, along with predictions based upon actual and effective depths applied to Equation \ref{betaEq}.  It can be seen that parameter $b$ is better predicted by the effective depth from Equation \ref{changeA} than the actual separation of the gratings. 

\subsection{Combining ASPs into a Fourier-complete array using CMOS}

\begin{figure} 
\centerline{\includegraphics[width=.6\textwidth]{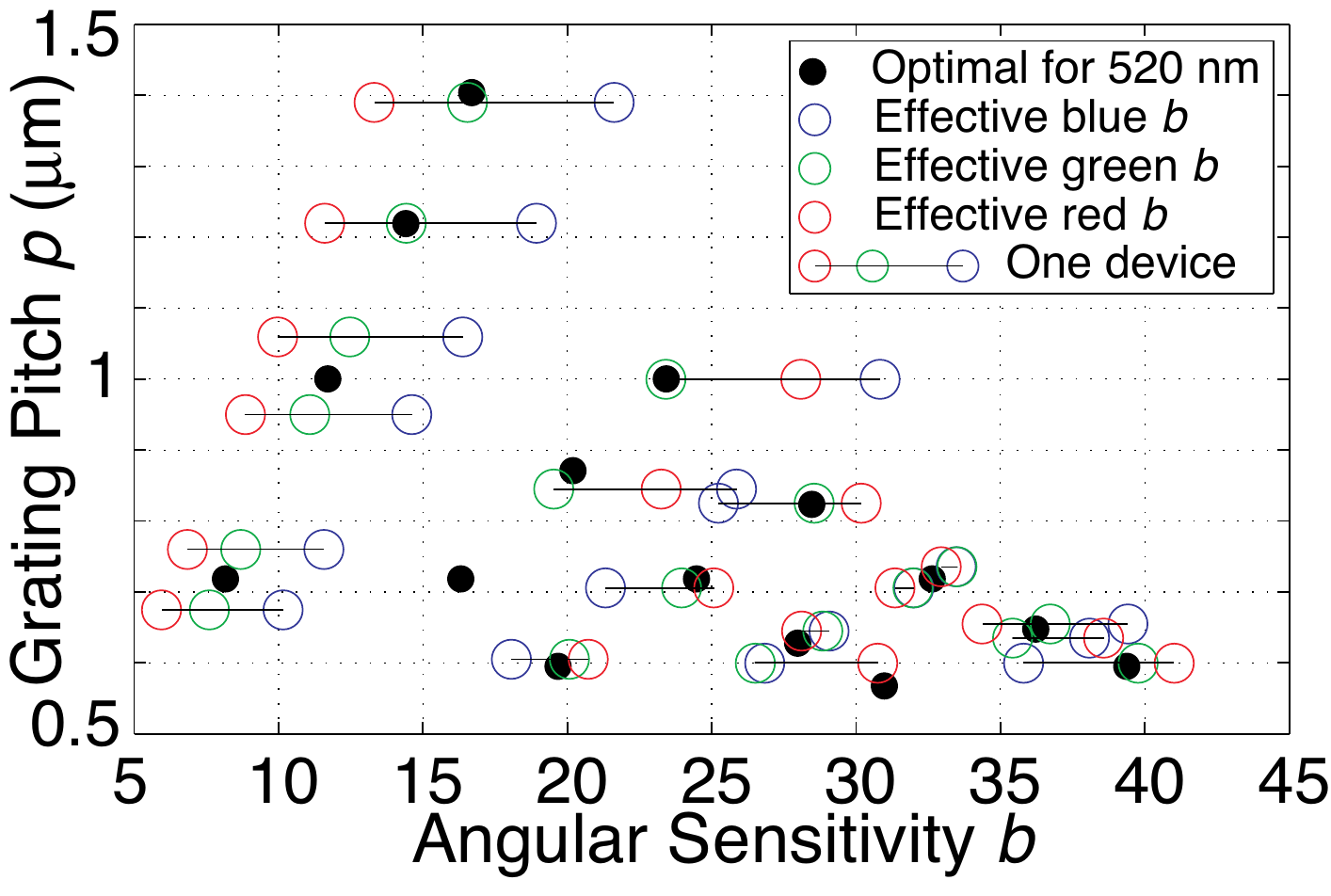}}
\caption{Selecting Devices for the PFCA.  Filled black circles indicate manufacturable devices with maximal $m$ for 520nm light; open circles indicate $p$ and ($\lambda$-dependent) $b$ for the suite of devices we manufactured.} 
\label{BPSelect4} 
\end{figure}

The metal interconnect layers of a CMOS process are manufactured at process-specific heights above the silicon substrate; to work within an established CMOS process it is necessary to manufacture all metals at one of the specified depths.  There are therefore a discrete spectrum of inter-metal depths available.  Using Equations \ref{talDep} and \ref{betaEq}, it is possible to determine the spectrum of manufacturable $b$ values that have locally-maximal $m$ (see filled circles of Figure \ref{BPSelect4}).  

For Fourier completeness, it is important to sample Fourier space with an adequate density to ensure coverage.  The more densely-packed the Fourier frequency measurements, the larger the angular region that can be observed with Nyquist sampling.  As reported previously \cite{PFCA2011}, the relationship between the maximum allowable incident light angle 
$h$ and the maximum difference in $b$ between consecutive designs is
\begin{equation}
  \label{maxEx}
h = \frac{180 \, ^{\circ}}{\sqrt{2} \Delta b}.
  \end{equation}

\begin{figure}[htb]
\centerline{\includegraphics[width=.6\textwidth]{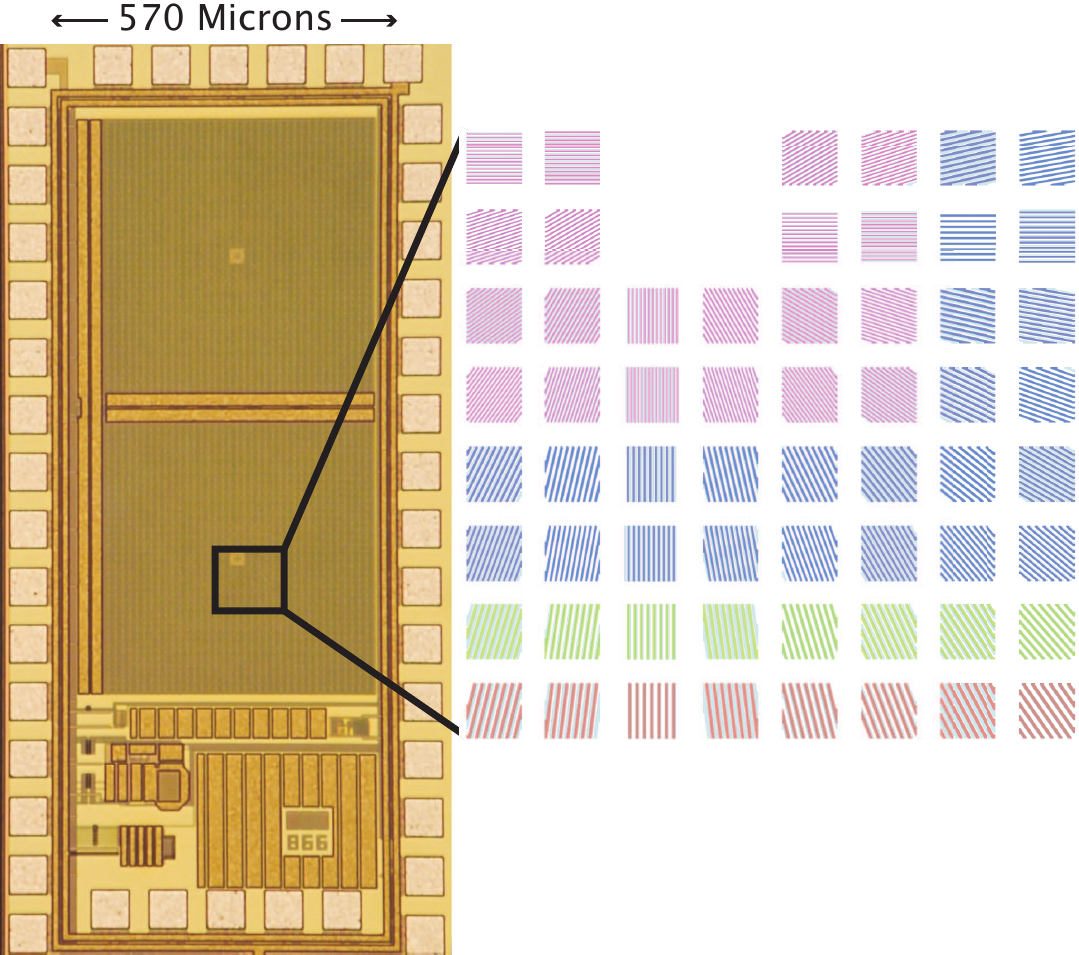}}
\caption{Manufactured PFCA.  Concentric rings of ASPs with increasingly higher $b$ yield a complete Fourier description of the light intensity from the far field.  Slowly varying orientation is evident from the enlarged section, where schematics show different metal layers in different colours.}
\label{fig_device}
\end{figure}

The first prototype we built included devices with $p$ tuned for a range of wavelengths other than the design point (520 nm) by manufacturing devices with $p$ differing slightly from that of locally-optimal devices (see Equation \ref{talDep}).  The ASPs included in the manufactured device are shown as empty circles in Figure \ref{BPSelect4}; note each design of constant $p$ and $z$ has different $b$ for different $\lambda$s.  We arranged 1444 ASPs in two PFCAs with complementary $\alpha$s (see Equation \ref{lightPix}).  All ASPs of a given design (combination of $z$ and $b$) are found in one of 18 concentric rings arranged around four low-$b$ devices described elsewhere  \cite{4550641}.  Rings with higher $b$ are placed further from the center to allow for a greater variety of grating orientations, as is required for Fourier tiling.  Although in general higher $b$ requires lower $p$, as seen in Figure \ref{BPSelect4} the first 6 devices chosen have the opposite trend since they are all designed to the first half-Talbot depth for four different metal layers.  The schematics in Figures \ref{fig_device} and \ref{fig_revcor} show $p$ increasing as one moves outward from the center, but in general this is not the case and the outermost device rings have the smallest $p$ and highest $b$.

Resolution limits are set by the maximum $b$ of any device $b_{max}$.  This Nyquist limit corresponds to two rows or columns of pixels per period of the highest-$b$ device.  For $h \approx 45 \, ^{\circ}$, the total number of effective pixels is $\left(\frac{b_{max}}{2}\right)^2$, or approximately 400 pixels for this prototype\footnote{As ASPs' transfer functions are not pure Fourier components but contain some harmonics, a limited amount of information outside the expected Nyquist limit is available so long as the coverage of Fourier space is overcomplete and noise levels are low.}.

The prototype's $b$s span the space relatively well for blue and green light (largest $\Delta b$ is 3.6 and 3.3 respectively; corresponding $h$s are $35  \, ^{\circ}$ and $38  \, ^{\circ}$), but a gap in $b$s for red light in the $b$ = 13--21 range reduces our expectations that red images will be recovered well for half-angles $h > 17  \, ^{\circ}$.  
Furthermore, as the pitch of a device approaches a single wavelength, its signal-to-noise ratios degrades, such that red light will intrinsically provide worse performance in fine-pitch ASPs.
Both of these disadvantages working under red light contribute to poor reconstructions (see Section \ref{colourRecon}).  The manufactured device is seen in Figure \ref{fig_device}.

\subsection{Reconstructing Images}

\begin{figure}[htb]
\centerline{\includegraphics[width=.6\textwidth]{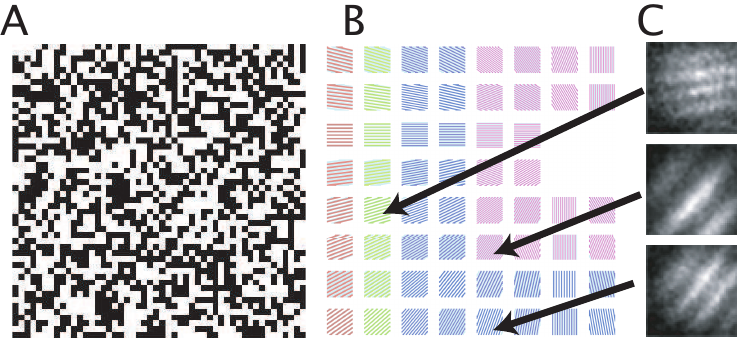}}
\caption{PFCA Calibration.  Transfer functions of each pixel are found by presenting 6710 random binary $50 \times 50$ calibration images (A) on a CRT screen (presentation time 16.7 ms each) to the array (B) and performing reverse correlation between the observed photocurrent of each sensor and the image presented.  The kernels of three ASPs are shown (C); these resemble Fourier components.}
\label{fig_revcor}
\end{figure}

The prototype PFCA was presented with calibration (Figure \ref{fig_revcor}A) and test images using a square CRT area 20cm on a side, 22.86cm from the PFCA\footnote{As 23cm $\gg$ 570$\mu$m (the PFCA's size), images presented are in the far field regime where the light field at each point in the PFCA is essentially identical.} for an $h$ of $31.7^{\circ}$ at the square's corners.  This $h$ is small enough to allow full Fourier coverage for blue and green, but not red light (see Equation \ref{maxEx}).

To reconstruct images computationally, we performed the following operations.  First, we computed the relationship between the photocurrent and the observed voltage drops at the photodiodes by fitting a quadratic function to $V(t)$ under constant illumination from an LED lit by a steady power source.  Then we threw out all data from any ASP that was saturated for more than 30\% of all image presentations (24.5\% of all sensors).  This was necessary because long range diffusion of photo-generated carriers tended to add a background photocurrent to many of the ASPs close to the edge of the array.  Next, we performed reverse correlation  \cite{theunissen2000spectral} between inferred photocurrents and the calibration images with the following steps.  We computed the pseudoinverse of the matrix\footnote{Matrix quantities such as $\mbf{C}$ shall be in capital bold, vector quantities in lowercase bold.} of calibration stimuli $\mbf{C}$, and multiplied it by the matrix of responses $\mbf{R}$ to obtain an estimate of the individual transfer functions $\mbf{H}$ as follows.  We assume $\mbf{R}$ is linearly related to $\mbf{C}$ by $\mbf{H}$, therefore
\begin{eqnarray}
\label{notherone}
\mbf{R} = \mbf{C H}. 
\end{eqnarray}

$\mbf{C}$ is not square since there are 6710 patterns of 2500 free parameters each, so we can compute and multiply both sides of Equation \ref{notherone} by $\mbf{C}$'s pseudoinverse $(\mbf{C^T C}) ^{-1} \mbf{C^T}$.
\begin{eqnarray}
(\mbf{C^T C}) ^{-1} \mbf{C^T R} & = & (\mbf{C^T C}) ^{-1} (\mbf{C^T C}) \mbf{H}. \nonumber \\
(\mbf{C^T C}) ^{-1} \mbf{C^T R} & = & \mbf{H}.\label{unbiased}
\end{eqnarray}

\begin{figure}[htb]
\centerline{\includegraphics[width=.6\textwidth]{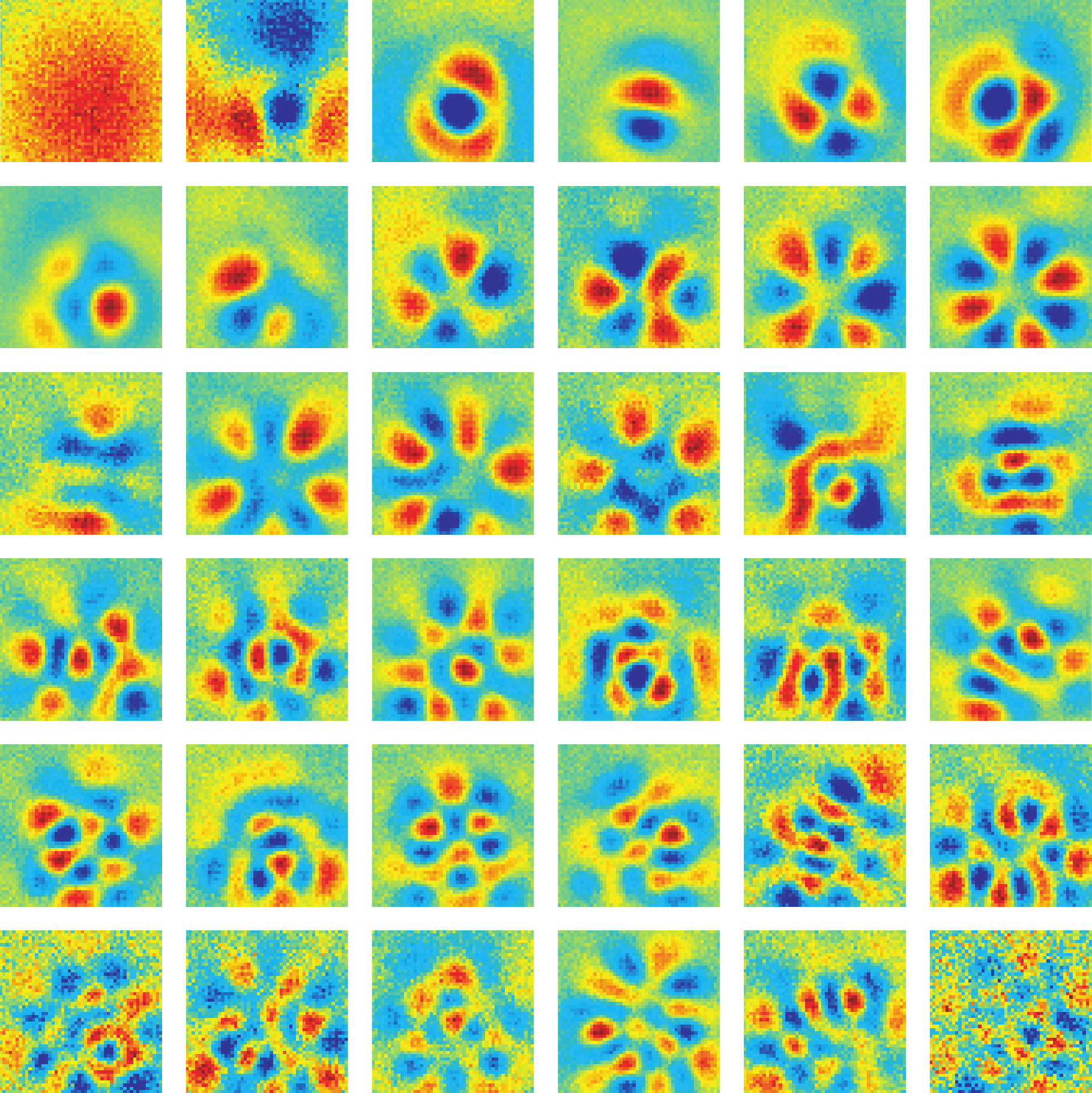}}
\caption{Eigenspectrum of $\mbf{H}$.  The strongest 36 eigenvectors of $\mbf{H}$ for green light are shown in order from left to right, then top to bottom.}
\label{36eigsnotext}
\end{figure}

Having computed $\mbf{H}$ (sample rows of $\mbf{H}$ are shown in Figure \ref{fig_revcor}C), we need to find a way to invert it so we can reconstruct a new stimulus $\mbf{s_{new}}$ given a new set of responses $\mbf{r_{new}}$.  We computed the eigenspectrum of $\mbf{H H^T}$ (see Figure \ref{36eigsnotext}) and found a regularized pseudoinverse of $\mbf{H}$ using ridge regression  \cite{ridge1943} as follows.  Ridge regression de-emphasizes eigenvectors of $\mbf{H H^T}$ that have little power.  It is an acknowledgement that some stimulus eigenvalues are not represented strongly in $\mbf{R}$ and thus should be attenuated to avoid letting noise in poorly-determined stimulus components overwhelm signal in well-determined components.\footnote{We also attempted ridge regression while determining $\mbf{H}$ in Equation \protect \ref{unbiased}, but it did not improve our results.}  

\begin{eqnarray}
\mbf{r_{new}} & = &  \mbf{s_{new} H}  \nonumber \\
\mbf{r_{new} H^T} (\mbf{H H^T} + \alpha \mathbb{I})^{-1} & \approx & \mbf{s_{new}} \label{recEq}
\end{eqnarray}

\begin{figure}[htb]
\centerline{\includegraphics[width=.6\textwidth]{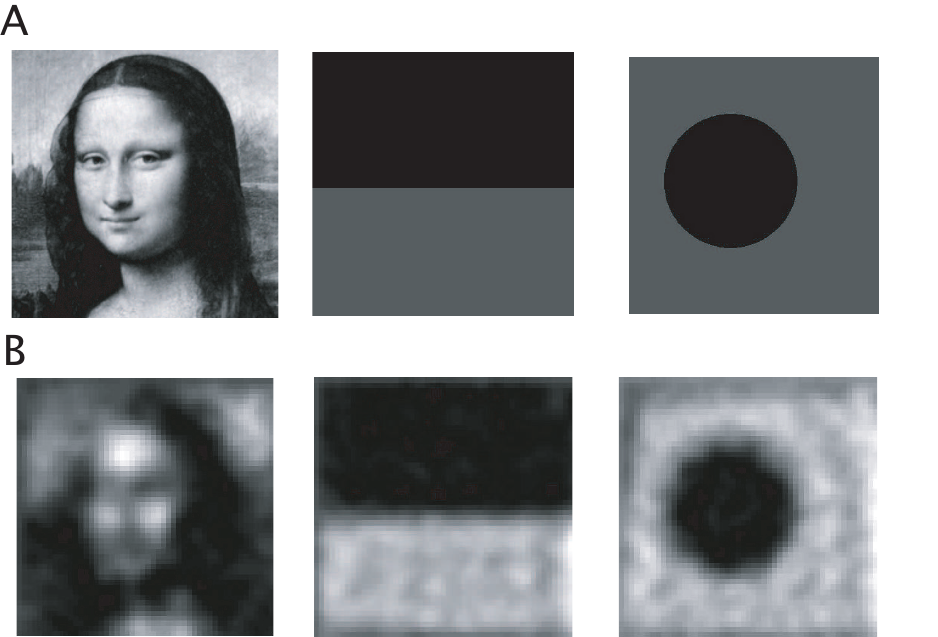}}
\caption{Image Reconstructions.  Using the basis functions obtained in the calibration phase (Fig. \protect \ref{fig_revcor}), we reconstructed (B) the image presented (A) up to the Nyquist limit of our array.  No off-chip optics were used; accumulation times were 16.7 ms.}
\label{fig_recon}
\end{figure}

Here $\alpha$ was chosen by inspection of the eigenspectrum of $\mbf{H H^T}$, and values of 30, 20 and 20 were used for reconstructions with red, green and blue light, respectively (see Section \ref{colourRecon}) to reflect cutoffs below which noise dominates signal.  These cutoffs preserved approximately 400 eigenvectors of $\mbf{H H^T}$ for green light, implying a $20 \times 20$ pixel resolution consistent with $\left(\frac{b_{max}}{2}\right)^2$.  Using Equation \ref{recEq}, we reconstructed presented images up to this resolution limit as in Figure \ref{fig_recon}.

\section{Robustness to Changes in Imaged Wavelength}
\label{colourRecon}
\begin{figure}[htb]
\centerline{\includegraphics[width=.6\textwidth]{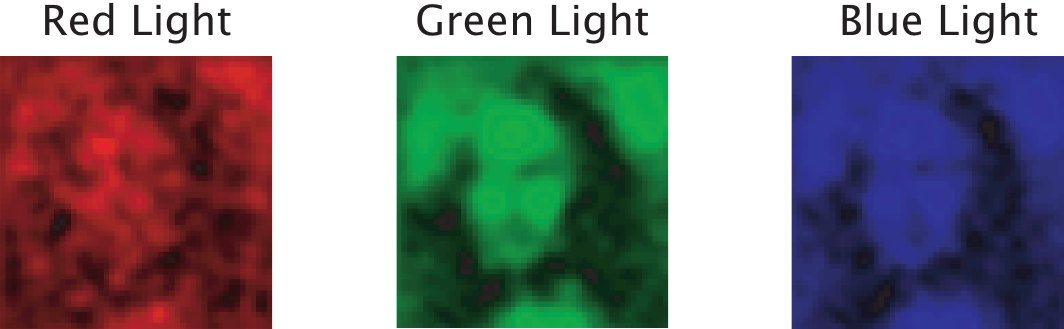}}
\caption{Three Colour Image Reconstructions.  Using the basis functions $\mbf{H}$ obtained from red, green and blue calibration runs, the first prototype chip is able to resolve light images other than at the design wavelength (520nm).  The center green image is identical to that of Figure \protect \ref{fig_recon}B on the left.}
\label{3colourRecon}
\end{figure}

Although the prototype PFCA was designed for green light, we also calibrated and tested it with blue and red light from the CRT.  For each wavelength of light, a separate $\mbf{H}$ was calculated (see Equation \ref{unbiased}) and images were reconstructed as for green light (see Equation \ref{recEq}). The reconstructed images for all three colours can be found in Figure \ref{3colourRecon}.

As expected, blue light does not produce images of the same quality as green light, and red light yields much worse reconstructions.  The poor reconstructions available from red light are due to the large gap in Fourier information available in the $b$ = 13--21 range\footnote{The PFCA's blind spot to frequencies in this range is more evident in Figure \protect \ref{blindRecon}, columns A1 and A2, rows G and B.  These images are derived from Equation \protect \ref{recEq} using the $\mbf{H}$ for red light but with green and blue light incident, leading to high noise.  Noise in the $b$ = 13--21 range is especially prominent, leading to artifactual ``bull's eyes'' in the reconstruction.} (see Figure \ref{BPSelect4}) and the fact that red light cannot pass easily through the smallest gratings in use due to its larger wavelength resulting in lower signal to noise ratios.

\section{Compressed Sensing Reconstructions}

\subsection{Brief Introduction to Compressed Sensing}

Compressed sensing (CS) is a recently-developed signal processing tool useful for reconstructing signals with only partial observations  \cite{1614066}.  
In general, one can only reconstruct a signal with as many (or fewer) free parameters as observations.  Partial observations imply a degraded reconstruction.  However, if the signal is known to be sparse in some basis, then incomplete observations in some other basis (unrelated to the basis in which the signal is sparse) can be sufficient to determine the signal with high precision.  For example, natural scenes are sparse in the space of wavelets \cite{olshausen1997sparse} (which, roughly speaking, encode edge information) since most scenes contain far fewer edges than pixels.  Furthermore, since most scenes are well described by their edges, one can encode most of the information about the pixels in a scene by encoding the strength of the edges in the scene.  Thus, a typical natural scene, when mapped onto a wavelet basis set, can be well approximated by a sparse collection of wavelets with large coefficients.  In general, if a system with $N$ free parameters (pixels) can be accurately described with $k$ (the number of edges) large-magnitude components, and the system is sparse ($k \ll N$), then accurate reconstruction can be performed from only $M$ measurements, even if $M<N$, provided $M>k$ by a comfortable\footnote{See \protect  \cite{1614066} for a theoretical guarantee of what ``comfortable'' means in this context, although provable guarantees place much more stringent requirements than practically necessary.} margin. This is true because only $k$ free parameters need to be fit.  Thus, by enforcing sparsity as well as accuracy, one can reconstruct many signals with fewer measurements than there are free parameters.

Compressed sensing provides a mathematical framework for arriving at a sparse solution that still explains observations using convex optimization\footnote{That the problem be posed as convex optimization is important because it eliminates the possibility of local optima, meaning the task of searching for a sparse signal compatible with incoming data is tractable.}.   CS operates first by assuming that data \mbf{y} form a $M \times 1$ vector derived from the $N \times 1$ sparse signal \mbf{x} multiplied by a $M \times N$ calibration matrix \mbf{A}.  The goal, then, is to find a signal $\mbf{\tilde{x}}$ that accurately estimates \mbf{x} based on \mbf{y}.  To make $\mbf{A\tilde{x}}$ close to \mbf{y}, minimizing the mean square error $\|\mbf{y}-\mbf{A\tilde{x}}\|^2_2$ is desirable.  To enforce sparsity, the $L_1$ norm of \mbf{x} $\|\mbf{\tilde{x}}\|_1$ should also be minimized \cite{1614066}.   Combining these two minimizations with relative importance $\lambda$ yields the basis pursuit denoising (BPDN) problem \cite{incrowd}:


\begin{eqnarray}
\label{BP}
\tilde{\mbf{x}} = \underset{\mrm{x}}{\operatorname{argmin}}\,\frac{1}{2}\|\mbf{y}-\mbf{Ax}\|^2_2+\lambda\|\mbf{x}\|_1
\end{eqnarray}
Solving Equation \ref{BP} leads to a sparse explanation of the signal, which will coincide with the true signal if it is indeed sparse.
\subsection{Compressed Sensing Permits Full Reconstructions from Incomplete PFCA Information}
\label{CSsec}
\begin{figure}[htb]
\centerline{\includegraphics[width=.6\textwidth]{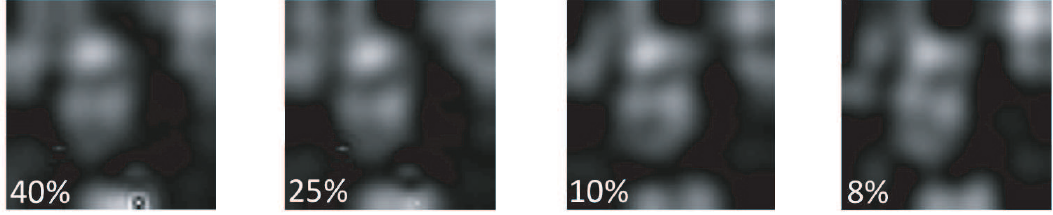}}
\caption{Compressed Sensing Results.  We show compressed sensing reconstructions of the test image using a randomly-selected subset of ASP sensors.  Proportion of sensors whose data are used shown in lower-left corner.}
\label{CSmona}
\end{figure}

The basis in which natural images are sparse is close to a wavelet pyramid  \cite{olshausen1997sparse}, which is different enough from the Fourier basis that a sparse collection of ASPs from a Fourier pixel should be sufficient to reconstruct a full image, provided it is sparse in the wavelet domain.  We therefore projected $\mbf{H}$ into an overcomplete ($2500 \times 8619$) wavelet basis  \cite{eero} $\mbf{W}$ and declared $\mbf{A}$ from Equation \ref{BP} to be $\mbf{HW}$.  The CS reconstruction problem for recovering the image given limited observations becomes 
\begin{eqnarray}
\label{BPW}
\mbf{s_{CS}} = (\mbf{W^T W})^{-1} \mbf{W^T} \left(\underset{\mrm{x}}{\operatorname{argmin}}\,\frac{1}{2}\|\mbf{r_{sub}}-\mbf{H_{sub}Wx}\|^2_2+\lambda\|\mbf{x}\|_1 \right)
\end{eqnarray}
where $\mbf{_{sub}}$ indicates only a randomly-chosen small subset (40\%, 25\%, 10\% or 8\%) of the responses $\mbf{r}$ are kept to simulate incomplete measurements.  CS reconstructions are of relatively high quality despite randomly subsampling Fourier space, as seen in Figure \ref{CSmona}.

As part of related work solving underdetermined imaging problems \cite{Sensors2010Poster}, we wrote the in-crowd algorithm for solving BPDN  \cite{incrowd}, which is faster than all alternative methods on sparse, large-scale BPDN problems such as this one. On a modern desktop computer, computing the exact BPDN solution for our problems takes approximately one second with the in-crowd algorithm, whereas other popular alternatives such as GPSR may take longer than an hour with the same inputs.  

PFCAs are thus able to relate complex images to a certain degree even when over 90\% of their constituent sensors malfunction.  Alternatively, it would be possible to design a PFCA with a random collection of only 10\% of the sensors traditionally considered essential for Fourier coverage and still be able to reconstruct images, provided they are sparse in some known, non-Fourier basis.  This ability is derived from the fact that each ASP makes measurements in a basis in which natural images are not sparse (i.e. Fourier components), yet natural images have a known sparsity.  There is only one sparse combination of wavelets that satisfies the incomplete Fourier measurements taken by the PFCA; this is the solution $\mbf{s_{CS}}$ of Equation \ref{BPW}.

\section{Determining Colour Blindly}

Up until now, we have assumed the reconstruction process has access to information about the wavelength of the light imaged.  Introducing uncertainty in the wavelength imaged makes reconstruction underdetermined by a factor equal to the number of possible wavelengths of light, since the number of unknowns becomes the number of pixels times the number of  $\lambda$s. However, recovering full colour information may still be possible.  In Section \ref{CSsec} we showed that an undercomplete subset of ASP readings can be sufficient to reconstruct a full image.  Moreover, as shown in Figure \ref{BPSelect4}, the $b$ of every device is wavelength-sensitive, meaning that PFCA signals should be sensitive enough to $\lambda$ to discriminate colours.  Given this wavelength sensitivity along with the robustness to undercomplete measurements as shown in Section \ref{CSsec}, it should be possible to determine the colour of light incident on the PFCA merely by assuming the image is composed of a sparse combination of wavelets of monochromatic light.

To perform this blind chromatic reconstruction we first computed the 
matrix of the coupling of red, green and blue wavelets to the sensor ensemble: $\mbf{[H_{\operatorname{red}} \, H_{\operatorname{green}} \, H_{\operatorname{blue}}]W_3}$ where $\mbf{W_3}$ denotes a triple-tall matrix of wavelet coefficients.  Next, we found the combination of these wavelets (of whatever mixture of colours) that is simultaneously sparse and satisfies the observed ASP outputs $\mbf{r}$ by solving Equation \ref{BPBlind}.

\begin{eqnarray}
\label{BPBlind}
\mbf{s_{CS}} = (\mbf{W_3^T W_3})^{-1}\mbf{ W_3}^T \left( \underset{\mrm{x}}{\operatorname{argmin}}\,\frac{1}{2}\|\mbf{r}-\mbf{[H_{\operatorname{red}} \, H_{\operatorname{green}} \, H_{\operatorname{blue}}]W_3 x}\|^2_2+\lambda\|\mbf{x}\|_1 \right)
\end{eqnarray}

\begin{figure}[htb]
\centerline{\includegraphics[width=.6\textwidth]{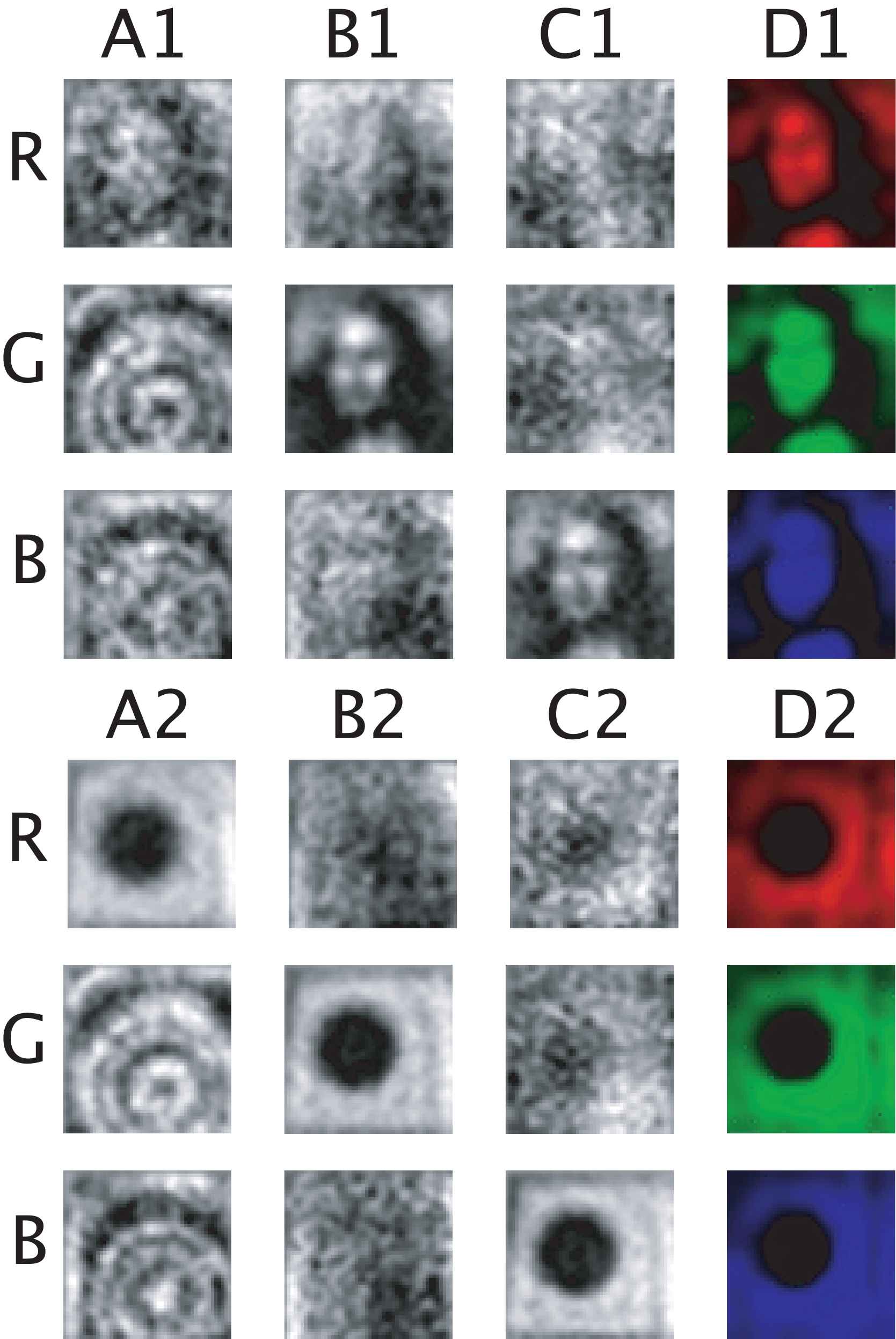}}
\caption{Blind Colour Imaging.  We show reconstructions of test images 1 \& 2 without foreknowledge of the colour of incident light.  R, G and B rows: presented  images  were red, green and blue.  A columns: image reconstructed using Equation \protect \ref{recEq} based on $\mbf{H}$ calibrated using red light.  B \& C columns: as for A, but using green- and blue-calibrated $\mbf{H}$ matrices.  D columns: reconstruction based on Equation \protect \ref{BPBlind} without any assumption about colour, or that the image is monochromatic.}
\label{blindRecon}
\end{figure}

The solution to Equation \ref{BPBlind} is the triple-tall red, green and blue vector of pixel intensities composed of a combination of a small number of monochromatic wavelets.  The solutions to Equation \ref{BPBlind} are plotted in Figure \ref{blindRecon}.  This technique does a remarkable job of discovering the correct colour.  Note that there is no constraint that all wavelets found by Equation \ref{BPBlind} must be of the same colour (and numerical inspection reveals that the images in D columns are not strictly monochromatic); merely enforcing sparsity of the solution and consistency with the observations selects wavelets that happen to be of the correct colour.  Note also that using compressed sensing greatly improves the performance of the PFCA for red light: compare R, D1 to R, A1 of Figure \ref{blindRecon}.  Unexpectedly, Fourier information from the missing band $b$ = 13--21 has been filled in.   CS gives no indication this completion should be possible since it prescribes  random sampling (with no systematic, large blind spots), yet it performs admirably despite a lack of theoretical guarantee.

\section{Conclusions}

We have demonstrated techniques for reconstructing images acquired by PFCAs that provide robustness beyond what was initially imagined possible.  First, we characterized wavelength robustness of our PFCA.  Although designed for green light, our first prototype PFCA performs well under blue light using standard linear algebra techniques (Figure \ref{3colourRecon}, right) and even under red light using basis pursuit denoising and compressed sensing (Figure \ref{blindRecon} R, D1 and R, D2).  Reconstruction with red incident light is particularly impressive given the large gap in spatial frequencies sampled under red light (Figure \ref{BPSelect4}) and the poor performance of ASPs whose $p$ approaches $\lambda$.  Second, we demonstrated robustness against losing all  signal from a randomly-chosen set of ASPs.  Using compressed sensing, we formulated a basis pursuit denoising problem (Equation \ref{BPW}) that permits impressive reconstructions even with a randomly-chosen set with 90\% of the signals destroyed (Figure \ref{CSmona}).  Last, we applied the same compressive sensing techniques to determine the colour of the images presented.  The reconstructed wavelengths are nearly flawless (Figure \ref{blindRecon}), and in the case of red illumination this technique yields an image of much higher quality than permitted by linear algebra reconstructions (Figure \ref{3colourRecon}, left).  In principle there is no obstacle to taking polychromatic images using a PFCA since Equation \ref{BPBlind} does not require images to be monochromatic, and it should be possible as well to recover hyperspectral information (i.e. more than 3 colour channels) in a compressed sensing framework, especially using a PFCA designed with a greater number of device types.  Each device type has an idiosyncratic $b(\lambda)$ with sharp discontinuities caused by abruptly rounding effective depth to the nearest optimal Talbot depth (see Figure \ref{placeholder} and Equation \ref{changeA}), providing much better $\lambda$ sensitivity than initially thought.

In summary, PFCAs enjoy several unusual forms of robustness stemming from their operating principles, including insensitivity to lost ASP sensors, flexibility in terms of incident light wavelength, and the ability to determine source image colour without employing any chromatic filters.  These unexpected forms of redundancy enhance the range of applications appropriate for this new class of sensing device.

\acknowledgments

We would like to thank Eve De~Rosa who helped support this work, Albert Wang for his input and ideas, and Igor Carron, whose suggestions and advice regarding compressed sensing and wavelet dictionaries proved invaluable.  We would also like to acknowledge our funding sources, including DARPA, who supported this research via a YFA Grant 66001-10-1-4028 to Alyosha Molnar, and the NIH, who helped fund this work under R21 grant EB 009841-01.


\begin{thebibliography}{9}

\bibitem{PFCA2011}
P.~R.~Gill, C.~Lee, D.~Lee, A.~Wang, and A.~Molnar, \emph{A microscale camera using direct Fourier-domain scene capture}, {Optics Letters}, vol.~36, no.~15, pp. 2949--2951, 2011.

\bibitem{wang2009light}
A.~Wang, P.~Gill, and A.~Molnar, \emph{Light field image sensors based on the
  Talbot effect}, {Applied optics}, vol.~48, no.~31, pp. 5897--5905,
  2009.

\bibitem{Sensors2010Poster}
A.~Wang, P.~R. Gill, and A.~Molnar, \emph{Fluorescent imaging and localization with
  angle sensitive pixel arrays in standard cmos}, presented at the IEEE
  Sensors Conference, 2010.
  
\bibitem{batchelor1985automated}
B.~Batchelor, D.~Hill, and H.~Hodgson, \emph{{Automated visual
  inspection}}.\hskip 1em plus 0.5em minus 0.4em\relax Elsevier Science Pub.,
  New York, NY, 1985.

\bibitem{lillesand2004remote}
T.~Lillesand, R.~Kiefer, and J.~Chipman, \emph{Remote sensing and image
  interpretation.}\hskip 1em plus 0.5em minus 0.4em\relax John Wiley \& Sons
  Ltd, 2004, no. Ed. 5.

\bibitem{urmson2008autonomous}
C.~Urmson, J.~Anhalt, D.~Bagnell, C.~Baker, R.~Bittner, M.~Clark, J.~Dolan,
  D.~Duggins, T.~Galatali, C.~Geyer \emph{et~al.}, \emph{Autonomous driving in
  urban environments: Boss and the urban challenge,} {Journal of Field
  Robotics}, vol.~25, no.~8, pp. 425--466, 2008.

\bibitem{bergstrom2009integration}
N.~Bergstr$\ddot{\mathrm{o}}$m, J.~Bohg, and D.~Kragic, \emph{Integration of visual
  cues for robotic grasping,} {Computer Vision Systems}, pp. 245--254,
  2009.
  
\bibitem{talbot1836lxxvi}
H.~Talbot, \emph{Lxxvi. facts relating to optical science. no. iv,}
  {Philosophical Magazine Series 3}, vol.~9, no.~56, pp. 401--407, 1836.

\bibitem{teng2008quasi}
S.~Teng, Y.~Tan, and C.~Cheng, \emph{Quasi-talbot effect of the high-density
  grating in near field,} {JOSA A}, vol.~25, no.~12, pp.~2945--2951,
  2008.
  
\bibitem{4550641}
C.~Koch, J.~Oehm, J.~Emde, and W.~Budde, \emph{Light source position measurement
  technique applicable in soi technology,} {Solid-State Circuits, IEEE
  Journal of}, vol.~43, no.~7, pp. 1588--1593, 2008.

\bibitem{theunissen2000spectral}
F.~E.~Theunissen, K.~Sen and A.~J.~Doupe, \emph{Spectral-temporal receptive fields of nonlinear auditory neurons obtained using natural sounds,} {Journal of Neuroscience},
  vol.~20, no.~6, pp.~2315-2331,2000.


\bibitem{ridge1943}
A.~N.~Tychonoff, \emph{On the stability of inverse problems,} {Doklady Akademii SSSR}, vol.~39, no.~5, pp.195--198, 1943.

%
\bibitem{1614066}
D.~Donoho, \emph{Compressed sensing,} {Information Theory, IEEE Transactions
  on}, vol.~52, no.~4, pp. 1289 --1306, 2006.

\bibitem{olshausen1997sparse}
B.~Olshausen and D.~Field, \emph{Sparse coding with an overcomplete basis set: A
  strategy employed by V1?} {Vision research}, vol.~37, no.~23, pp.
  3311--3325, 1997.
  
\bibitem{incrowd}
P.~R.~Gill, A.~Wang, and A.~Molnar, \emph{The in-crowd algorithm for fast basis pursuit denoising,} {Signal Processing, IEEE Transactions on},vol.~59, no.~10, pp. 4595--4605, 2011.  Code available at \verb6http://molnargroup.ece.cornell.edu/files/InCrowdBeta1.zip6.

\bibitem{eero}
E.~P.~Simoncelli and W.~T.~Freeman, \emph{The steerable pyramid: a flexible architecture for multi-scale derivative computation,} {IEEE Second Int'l Conf. on Image Processing}, Washington DC, October 1995.  Code available at \verb6http://www.cns.nyu.edu/~eero/steerpyr/6.


%
%
%

\end{thebibliography}
\end{document}